\documentclass[]{spie}

\usepackage{amsmath,amsfonts,amssymb}
\usepackage{graphicx}
\usepackage[colorlinks=true, allcolors=blue]{hyperref}
\title{Need for objective task-based evaluation of AI-based segmentation methods for quantitative PET}

\author[a]{Ziping Liu}
\author[b]{Joyce C. Mhlanga}
\author[b,c]{Barry A. Siegel}
\author[a,b,c]{Abhinav K. Jha}
\affil[a]{Department of Biomedical Engineering, Washington University in St. Louis, St. Louis, MO, USA}
\affil[b]{Mallinckrodt Institute of Radiology, Washington University School of Medicine, St. Louis, MO, USA}
\affil[c]{Alvin J. Siteman Cancer Center, Washington University School of Medicine, St. Louis, MO, USA}

\authorinfo{Corresponding author: Abhinav K. Jha (E-mail: a.jha@wustl.edu)}

\begin{document}

\begin{titlepage}

This manuscript has been accepted to SPIE Medical Imaging, February 19-23, 2023. Please use the following reference when citing the manuscript.

Liu, Z., Mhlanga, J. C., Siegel, B. A., and Jha, A. K., ``Need for objective task-based evaluation of AI-based segmentation methods for quantitative PET", Proc. SPIE Medical Imaging, 2023.

\end{titlepage}

\maketitle

\begin{abstract}
Artificial intelligence (AI)-based methods are showing substantial promise in segmenting oncologic positron emission tomography (PET) images. 
For clinical translation of these methods, assessing their performance on clinically relevant tasks is important. 
However, these methods are typically evaluated using metrics that may not correlate with the task performance.
One such widely used metric is the Dice score, a figure of merit that measures the spatial overlap between the estimated segmentation and a reference standard (e.g., manual segmentation). 
In this work, we investigated whether evaluating AI-based segmentation methods using Dice scores yields a similar interpretation as evaluation on the clinical tasks of quantifying metabolic tumor volume (MTV) and total lesion glycolysis (TLG) of primary tumor from PET images of patients with non-small cell lung cancer. 
The investigation was conducted via a retrospective analysis with the ECOG-ACRIN 6668/RTOG 0235 multi-center clinical trial data. 
Specifically, we evaluated different structures of a commonly used AI-based segmentation method using both Dice scores and the accuracy in quantifying MTV/TLG. 
Our results show that evaluation using Dice scores can lead to findings that are inconsistent with evaluation using the task-based figure of merit. 
Thus, our study motivates the need for objective task-based evaluation of AI-based segmentation methods for quantitative PET. 
\end{abstract}


\section{INTRODUCTION}
\label{sec:intro}
Quantitative positron emission tomography (PET) is showing substantial promise in multiple clinical applications. 
A major area of interest is in evaluating tumor volumetric features, including metabolic tumor volume (MTV) and total lesion glycolysis (TLG), as prognostic and predictive biomarkers \cite{chen2012prognostic,ohri2015pretreatment}.
Similarly, there is substantial interest in evaluating PET-derived radiomic features (e.g., intra-tumor heterogeneity) as biomarkers for predicting cancer treatment outcomes \cite{mena201718f,lee2018radiomics}.
In all these applications, accurate segmentation of tumors on PET images is required.
Given this clinical importance, multiple segmentation methods are being actively developed for oncologic PET.
In particular, artificial intelligence (AI)-based segmentation methods have been showing substantial promise in yielding tumor delineations on oncologic PET images \cite{zhao2018tumor,leung2020physics,liu2021bayesian,yousefirizi2021toward}.
For clinical translation of these methods, appropriate evaluation procedures are required.

Medical images are acquired for a specified clinical task.
Thus, it is widely recognized that imaging systems and algorithms be evaluated on their performance in the clinical task.
In this context, strategies for performing task-based assessment of image quality have been developed \cite{barrett1995objective,barrett1998objective,barrett2015task,jha2021objective}.
However, AI-based algorithms are often evaluated using metrics that may not account for the clinical task of interest.
This can be an issue since evaluation using these task-agnostic metrics may yield interpretations that are not consistent with evaluation on the clinical task.
For example, Yu et al. \cite{yu2023need} observed that evaluation of an AI-based denoising method using conventional fidelity-based metrics such as the structural similarity index measure (SSIM) led to findings that were inconsistent with evaluation on the clinical task of detecting myocardial perfusion defects.
Similar findings have been observed in multiple other studies \cite{badal2019virtual,li2021assessing,prabhat2021deep}, all motivating the need for task-based evaluation of AI-based denoising algorithms.

Similar to the AI-based denoising methods, currently, AI-based segmentation methods are also typically evaluated using metrics that are not explicitly designed to measure performance on clinical tasks \cite{jha2021objective,liu2022tissue}.
These metrics quantify the similarity between the estimated segmentation and a reference standard (e.g., manual segmentation).
One such commonly used metric is the Dice score \cite{foster2014review}.
This metric measures the spatial overlap between the reference standard and estimated segmentation.
A higher value of the Dice score is often used to infer a more accurate performance.
Other similar metrics include the Hausdorff distance and Jaccard similarity coefficient.
However, as mentioned above, these metrics may not correlate with the task performance.
For example, on the clinical task of quantifying the MTV of primary tumor, it is unclear how the value of Dice score correlates with the accuracy in the quantified MTV.

Given the wide use of task-agnostic metrics to evaluate AI-based segmentation methods for oncologic PET, the goal of our work was to investigate whether such evaluation yields a similar inference to evaluation based on the performance in clinical tasks.
Towards this goal, we conducted a retrospective study with the ECOG-ACRIN 6668/RTOG 0235 multi-center clinical trial data \cite{machtay2013prediction,kinahan2019data}.
In this study, we compared the inferences obtained from evaluation of a commonly used AI-based segmentation method using the task-agnostic metric of Dice score vs. performance on the clinical tasks of quantifying the MTV and TLG of primary tumor from PET images of patients with stage IIB/III non-small cell lung cancer (NSCLC). 

\section{Methods}
\label{sec:methods}
In this section, we first present the procedure to collect the oncologic PET images and obtain segmentation annotations (sec.~\ref{sec:methods(data collection)}). 
We then describe the considered AI-based method to segment the primary tumor on those PET images in section~\ref{sec:methods(AI-based method)}. 
Finally, in sections~\ref{sec:methods(sub:evaluation)} and \ref{sec:methods(sub:FoM)}, we describe the procedure to evaluate this method based on the Dice score and task-based figure of merit.

\subsection{Data collection and curation}
\label{sec:methods(data collection)}
In this study, de-identified $^{18}$F-fluorodeoxyglucose (FDG)-PET and computed tomography (CT) images of 225 patients with stage IIB/III NSCLC were collected from The Cancer Imaging Archive (TCIA) database \cite{clark2013cancer}.
All the FDG-PET/CT images were obtained prior to chemoradiotherapy.
For each patient, a board-certified physician (J.C.M) with more than 10 years of experience in reading nuclear medicine scans segmented the primary tumor on the PET images.
Specifically, the physician first identified the primary tumor by reviewing the PET/CT images in axial, sagittal, and coronal planes using a vendor-neutral software (MIM Encore 6.9.3; MIM Software Inc., Cleveland, OH).
The physician then used an edge-detection tool provided by MIM Encore to obtain an initial boundary of the tumor.
This boundary, after being examined and adjusted by the physician, was used to obtain the eventual segmentation of the primary tumor.
In our evaluation study, these manual delineations were considered as a surrogate for ground truth for training and evaluating the considered AI-based segmentation method, as will be described in the next section. 

\subsection{Considered AI-based segmentation method}
\label{sec:methods(AI-based method)}

\subsubsection{Network architecture}
\label{sec:methods(sub:network architecture)}
\begin{figure}[htbp]
\centering
\includegraphics[width=\textwidth]{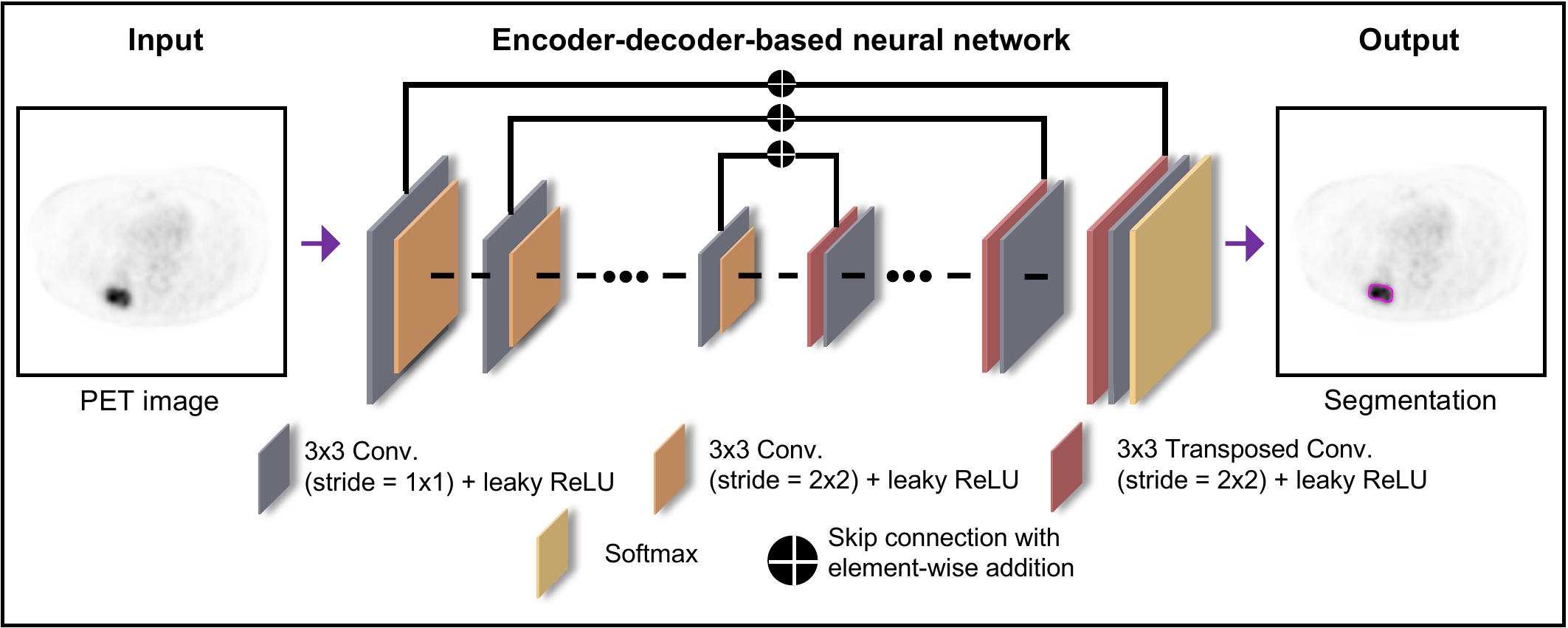}
\caption{Architecture of the considered commonly used U-net-based convolutional neural network model.}
\label{fig: CNN model}
\end{figure}

We considered a commonly used U-net-based convolutional neural network (CNN) model \cite{zhao2018tumor,leung2020physics,blanc2018automatic} to segment the primary tumor on 3-D PET images on a per-slice basis.
Specifically, given a PET image slice that contains the primary tumor, the CNN model segmented the tumor by performing voxel-wise classification (figure~\ref{fig: CNN model}).
This CNN model was partitioned into an encoder and a decoder. 
The encoder was designed to extract spatially local features of the input PET images through successive blocks of convolutional layers.
The decoder was then designed to use these extracted features to segment the primary tumor through blocks of transposed convolutional layers.
After each layer in the encoder and decoder, a leaky rectified linear unit activation function was applied.
Additionally, between the block of layers in the encoder and decoder, skip connections with element-wise addition were applied to stabilize the network training \cite{mao2016image}.
Further, dropout was applied in each layer to reduce network overfitting.
In the final layer of the CNN model, a softmax function was applied to the output of the decoder to perform the voxel-wise tumor classification.

\subsubsection{Network training}
\label{sec:methods(sub:network training)}
Of the total 225 patients, 180 were used for training the considered CNN model.
During training, 2-D PET images with the corresponding surrogate ground truth (i.e., manual segmentations obtained following the procedure described in section~\ref{sec:methods(data collection)}) were input to the network.
The network was then trained to minimize a cost function based on the binary cross-entropy (BCE) loss between the true and estimated segmentations, using the Adam optimization algorithm \cite{kingma2014adam}.
Network hyperparameters were optimized via five-fold cross-validation on the training data.
Data from the remaining 45 patients were reserved for evaluating the trained network, as will be described in the next section.

\subsection{Evaluation of the considered AI-based method}
\label{sec:methods(sub:evaluation)}
An important factor that is known to impact the performance of AI-based methods is the choice of network depth \cite{sun2016depth}.
Thus, we investigated whether evaluating the considered CNN model with different network depths based on the Dice scores yields a similar interpretation to evaluation based on clinical-task performance.
Specifically, we varied the depth of the CNN model by setting the paired number of blocks of layers in the encoder and decoder (section~\ref{sec:methods(sub:network architecture)}) to 2, 3, 4, and 5.
For each choice of depth, the network was independently trained and cross-validated on images of 180 patients (section~\ref{sec:methods(sub:network training)}).
The trained network was then evaluated on the 45 test patients based on both the task-agnostic metric of Dice scores and performance on quantifying the MTV/TLG of the primary tumor, as will be detailed in the next section.

\subsection{Figures of Merit for evaluation}
\label{sec:methods(sub:FoM)}
The performance of the considered CNN model was first quantified using the task-agnostic figure of merit of Dice score.
As introduced in section~\ref{sec:intro}, the Dice score measures the spatial overlap between the true and estimated tumor segmentations, denoted by $S_t$ and $S_e$, respectively.
Denote the overlap between $S_t$ and $S_e$ by $S_t \cap S_e$, with $|S_t \cap S_e|$ denoting the number of voxels that are correctly classified as belonging to the tumor class.
For each patient, the Dice score is given by
\begin{align}
    \textrm{Dice score}
    =
    \dfrac{2|S_t \cap S_e|}{|S_t| + |S_e|}.
\end{align}

We then evaluated the CNN model on the clinical tasks of quantifying the MTV and TLG of the primary tumor.
The accuracy in estimating the MTV/TLG was quantified based on absolute ensemble normalized bias (NB) between the MTV/TLG obtained with the manual segmentation (considered as ground truth) and those obtained with the segmentation yielded by the CNN model. 
Denote the number of patients by $P$.
For the $p^\mathrm{th}$ patient, denote the true and estimated MTV  by $V_p$ and $\hat{V}_p$, respectively.
Similarly, denote the true and estimated TLG of the $p^\mathrm{th}$ patient by $G_p$ and $\hat{G}_p$, respectively.
The absolute ensemble NB of the estimated MTV and TLG are given by
\begin{subequations}
\begin{align}
\textrm{Absolute ensemble NB (MTV)} = \left|\dfrac{1}{P} \sum_{p=1}^P \dfrac{\hat{V}_p - V_p}{V_p}\right|, \\
\textrm{Absolute ensemble NB (TLG)} = \left|\dfrac{1}{P} \sum_{p=1}^P \dfrac{\hat{G}_p - G_p}{G_p}\right|.
\end{align}
\end{subequations}
A lower value of absolute ensemble NB indicates a higher accuracy in the estimated MTV/TLG.

\section{Results}
\label{sec:results}

\begin{figure}[htbp]
\centering
\includegraphics[width=\textwidth]{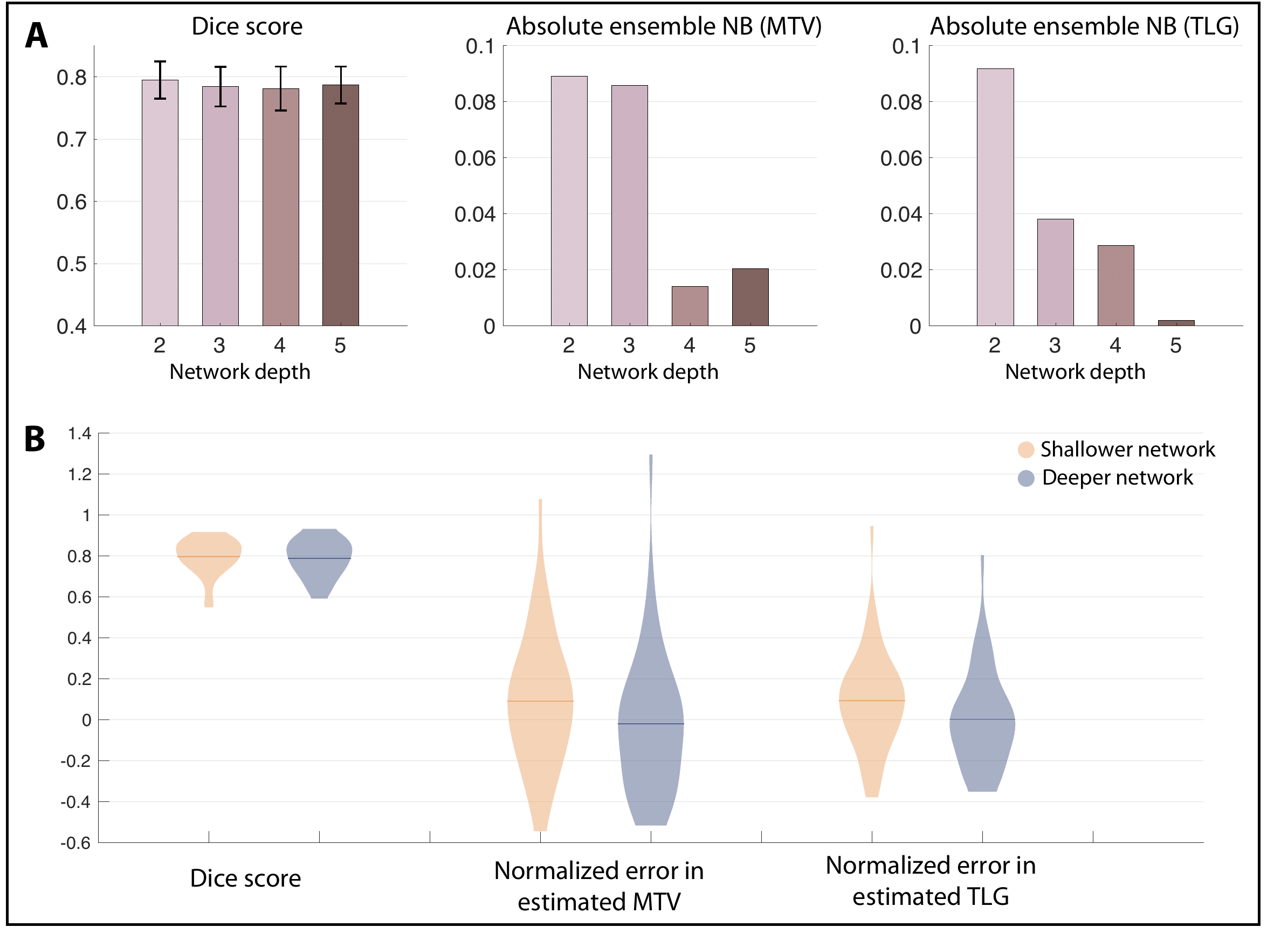}
\caption{Task-based evaluation of the impact of network depth on the performance of the considered CNN model. (A) Performance quantified using the task-agnostic Dice scores and task-based figure of merit of absolute ensemble normalized bias; (B) Violin plot showing the distributions of the Dice scores and normalized error in estimated MTV/TLG for a shallower and deeper network, respectively.}
\label{fig: task-agnostic vs. task-based}
\end{figure}

Figure~\ref{fig: task-agnostic vs. task-based}(A) shows the performance of the considered CNN model with different network depths, as quantified using both the task-agnostic metric of Dice scores and the task-based figure of merit of absolute ensemble NB.
Based on the Dice scores, we observe that there was no significant difference (p $<$ 0.01) between the different network depths.
However, deeper networks (with more than three paired blocks of convolutional layers) actually yielded substantially lower values of absolute ensemble NB in estimated MTV/TLG.

Figure~\ref{fig: task-agnostic vs. task-based}(B) shows the distributions of the Dice scores and normalized error in estimated MTV/TLG for a shallower and a deeper network. 
We observe that the deeper network yielded normalized errors that were more evenly distributed around zero and less positively skewed.
Consequently, this led to a lower value of absolute ensemble NB approaching zero.

\begin{figure}[htbp]
\centering
\includegraphics[width=\textwidth]{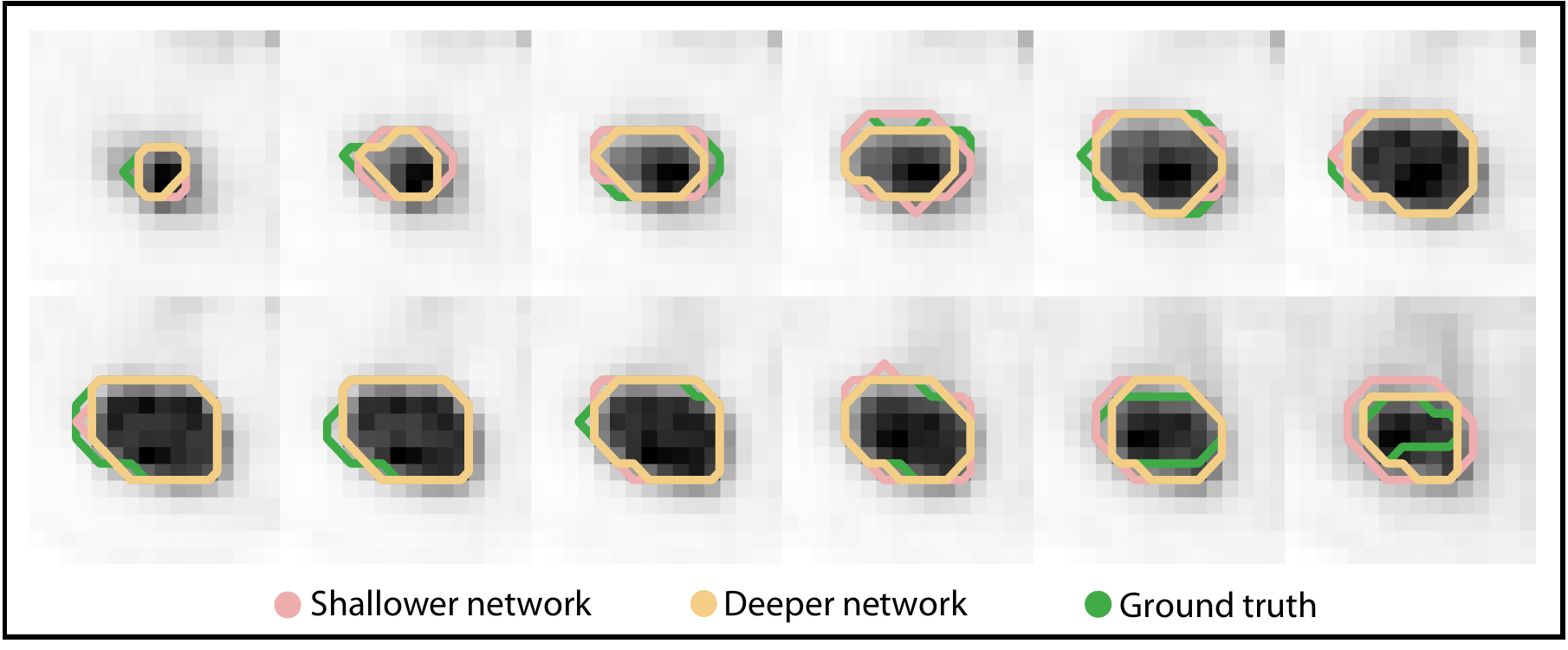}
\caption{Segmentations of the primary tumor in a representative patient obtained with a shallower and a deeper network. The segmentations for each of the slices that contained the tumor are shown. For this patient, the Dice scores were very similar between the shallower and deeper network, even though the errors in estimating MTV/TLG were very different. We also note that visually, the segmentations are different for the shallower vs. deeper networks.}
\label{fig: visual inspection}
\end{figure}

Figure~\ref{fig: visual inspection} shows the qualitative comparison of segmentations of the primary tumor in a representative test patient as yielded by the shallower and deeper network.
For this patient, the deeper network yielded very similar Dice scores compared to the shallower network (0.90 vs. 0.91).
However, the deeper network yielded substantially lower normalized error in both the estimated MTV (0.02 vs. 0.11) and TLG (0.03 vs. 0.09).

\section{Discussion and conclusion}
\label{sec:discussion}
AI-based segmentation methods are showing substantial promise in oncologic PET.
However, these methods have typically been evaluated using metrics that are task-agnostic, such as the Dice score. 
Since medical images are acquired for a specified clinical task, it is important that the AI-based methods be objectively evaluated using figures of merit that directly correlate with that task.
The key contribution of our study is to verify this important need by investigating whether evaluation of AI-based segmentation methods using the task-agnostic Dice scores yields similar interpretations to evaluation on the clinical tasks of quantifying the MTV/TLG of primary tumor from PET images of patients with NSCLC.

Our results from section~\ref{sec:results} indicate that evaluation of a commonly used AI-based segmentation method using the task-agnostic Dice score could lead to findings that were discordant with evaluation on the clinical tasks of MTV/TLG quantification.
Initially, we observe in figure~\ref{fig: task-agnostic vs. task-based}(A) that deeper networks yielded very similar Dice scores compared to shallower networks.
This could then lead to the inference that a deeper network did not improve the performance on segmenting the tumor.
Given the high demand for computational resources when training AI-based methods, one may prefer to deploy shallower networks in clinical studies from this initial inspection.
However, our results show that a deeper network actually demonstrated more accurate performance on the clinical tasks of MTV/TLG quantification by yielding substantially lower values of absolute ensemble NB.
These results indicate the limited ability of the Dice score to objectively evaluate AI-based segmentation methods, thus demonstrating the importance of task-based evaluation.

Evaluation of the effect of AI-based segmentation methods on quantitative tasks requires the knowledge of true quantitative values of interest.
However, such ground truth may typically be unavailable in clinical studies.
We circumvented this challenge by considering the manual segmentation as defined by the physician as the ground truth.
However, we recognize this as a limitation of this study.
To address this limitation, no-gold-standard evaluation techniques have been developed \cite{kupinski2002estimation,hoppin2002objective,jha2016no,liu2022no}.
These techniques have demonstrated the efficacy in evaluating PET segmentation methods on clinically relevant tasks even without any knowledge of the ground truth \cite{jha2017practical,liu2020no,zhu2021comparing}.
Thus, these techniques could provide a mechanism to perform objective task-based evaluation of segmentation methods in the absence of ground truth.
Another limitation of this study is that we evaluated the considered AI-based segmentation method only on the clinical tasks of quantifying MTV/TLG.
Expanding our evaluation for other tasks such as quantifying intra-tumor heterogeneity is an important future research direction.

The findings from our study further motivate the recently published Recommendations for Evaluation of AI for NuClear medicinE (RELAINCE) guideline \cite{jha2022nuclear} proposed by the SNMMI AI task force.
The task force advocated that while task-agnostic metrics such as the Dice score are valuable for assessing the initial promise of a segmentation method, it is important to further quantify the technical performance of the method on the clinical task for which imaging is performed.
Our results further confirmed the important need for this task-based evaluation.

In conclusion, the conducted retrospective analysis with the ECOG-ACRIN 6668/RTOG 0235 multi-center clinical trial data shows that evaluation of a commonly used AI-based segmentation method using the task-agnostic metric of Dice scores can lead to findings that are inconsistent with task-based evaluation. 
Thus, the results emphasize the need for objective task-based evaluation of AI-based segmentation methods for quantitative PET.

\section*{ACKNOWLEDGEMENTS}
Financial support for this work was provided by the National Institute of Biomedical Imaging and Bioengineering R01-EB031051, R01-EB031962, R56-EB028287, and R21-EB024647 (Trailblazer Award).

\bibliography{ref}
\bibliographystyle{spiebib}

\end{document}